%% file: paper.tex
\documentclass[preprintnumbers,preprint,nofootinbib,amsmath,amssymb]{revtex4}

\usepackage[dvips]{graphicx}
\usepackage{amsmath,amssymb}

\input{mydefines.tex}

\preprint{RESCEU-8/08}

\begin{document}

\title{Single-field inflation, anomalous enhancement of superhorizon fluctuations, and non-Gaussianity in primordial black hole formation}
\author{Ryo Saito$^{1,2}$} \author{Jun'ichi Yokoyama$^{2}$} \author{Ryo Nagata$^{2}$}
\affiliation{$^1$Department of Physics, Graduate School of Science,  
The University of Tokyo, Tokyo 113-0033, Japan\\
$^2$Research Center for the Early Universe (RESCEU),
Graduate School of Science, The University of Tokyo,Tokyo 113-0033, Japan
}

\begin{abstract}
 We show a text-book potential for single-field inflation, namely,
the Coleman-Weinberg model can induce double inflation and formation
of primordial black holes (PBHs), because fluctuations that leave
the horizon near the end of first inflation are anomalously enhanced
at the onset of second inflation when the time-dependent mode turns
to a growing mode rather than a decaying mode.  The mass of PBHs
produced in this mechanism lies in several discrete ranges depending
on the model parameters. We also calculate the effects of non-Gaussian
statistics due to higher-order interactions on the abundance of
PBHs, which turns out to be small.
\end{abstract}

\maketitle


 \section{Introduction}\label{sec:intro}
  The primordial perturbation generated in the inflationary epoch \cite{inflation, pert} is believed to be the origin of large-scale structure observed in the universe today. We have accurate information on the primordial perturbation through the observation of anisotropy of the cosmic microwave background(CMB) with the help of cosmological perturbation theory \cite{wmap5}. However, the CMB observation provides us with the information on the perturbations on a limited range of scales. Therefore, we cannot say, a priori, anything on the primordial perturbation at the smaller scales from CMB data. If the perturbation at these scales is order unity, primordial black holes can be produced when the scale of overdensed region crosses the horizon \cite{pbh}. The typical mass of these black holes is given by horizon mass at horizon crossing\footnote{Strictly speaking, the mass of black hole depends on the density perturbations $\delta$ and black holes with a small fraction of horizon mass are also produced \cite{niemeyer1999}. Since the contributions of these black holes are small compared to the black holes with horizon mass, we ignore this dependence for simplicity.}:
	\begin{equation}\label{eq:bhmass}
		M_{\text{BH}} = \frac{4\pi}{3}\rho(H^{-1})^3 = \frac{4\pi M_{G}^2}{H},
	\end{equation}	
where $M_G=(8\pi G)^{-1/2}=2.4\times10^{18}~\mr{GeV}$ is the reduced Planck mass\footnote{We use units for which $c=1$.}. Though PBH evaporates through the Hawking radiation process \cite{hawking}, those with mass greater than $10^{15}~\mr{g}$ can remain until the present time \cite{carr2003,carr1976}. These PBHs can be an origin of intermediate mass black holes \cite{kawaguchi} or dark matter in the Universe \cite{carr2003,ivanov}.

It is difficult to produce appreciable numbers of PBHs in the simple single-field slow-roll inflation models which predict nearly scale-invariant power spectrum of curvature perturbations because the amplitude of fluctuations on small scales cannot be much different from that normalized by CMB observation on large scales. One exception among
others \cite{models, chong2007} is the chaotic new inflation model \cite{yokoyama1998} where double inflation is realized with a single field. In Ref.~\cite{yokoyama1998}, however, calculation of curvature fluctuation was done using the slow-roll formula in which only the time-independent mode has been taken into account.  In the present paper, we solve the  evolution equation 
of each Fourier mode of fluctuations properly and find an anomalous growth due to the temporal deviation from slow-roll evolution between two inflationary stages.  This provides the first realistic example of the anomalous growth of perturbation in the superhorizon regime discussed in Ref.~\cite{leach2001}.  As a result we obtain a power spectrum highly peaked on some scales depending on the values of the model parameters and PBH formation is more easily realized than concluded in Ref.~\cite{yokoyama1998}. We search for the values of parameters with which appreciable numbers of PBHs are produced under the observational constrains \cite{constraints}. We also analyze the effects of non-Gaussianity generated in this model on the abundance of PBHs.

The organization of this paper is as follows. In section \ref{sec:model}, we introduce the chaotic new inflation model and explain the background evolution in this model. In section \ref{sec:enhance}, a mechanism of the enhancement of the perturbation is explained and the power spectrum in the chaotic new inflation model is given. In section \ref{sec:abundance}, we give an expression of PBH abundance resulted from a peaked power spectrum. In section \ref{sec:ng}, we estimate non-Gaussian correction to PBH abundance. In section \ref{sec:search}, we calculate the PBH abundance with various values of parameters and give a relation between mass and PBH abundance. Section \ref{sec:conclusion} contains conclusion of this paper. In this paper, we use curvature perturbation in the comoving gauge $\zeta$ as a degree of freedom of scalar perturbation.


 \section{Chaotic new inflation model}\label{sec:model}
~~We consider a single-field inflation model with the Coleman-Weinberg potential \cite{cw}
	\begin{equation}\label{eq:pot}
		V(\vphi) = \frac{\lambda}{4}\vphi^4\left( \ln \left|\frac{\vphi}{v}\right| - \frac{1}{4}\right) + \frac{\lambda}{16}v^4.
	\end{equation}
~~Historically, this potential was first used to realize new inflation \cite{new}, but phase space consideration has led to the conclusion that chaotic inflation \cite{chaotic} is much more likely to occur \cite{kung}. So we start with a large field value. 

\begin{figure}[h]
		\centering
		\includegraphics[width=.47\linewidth]{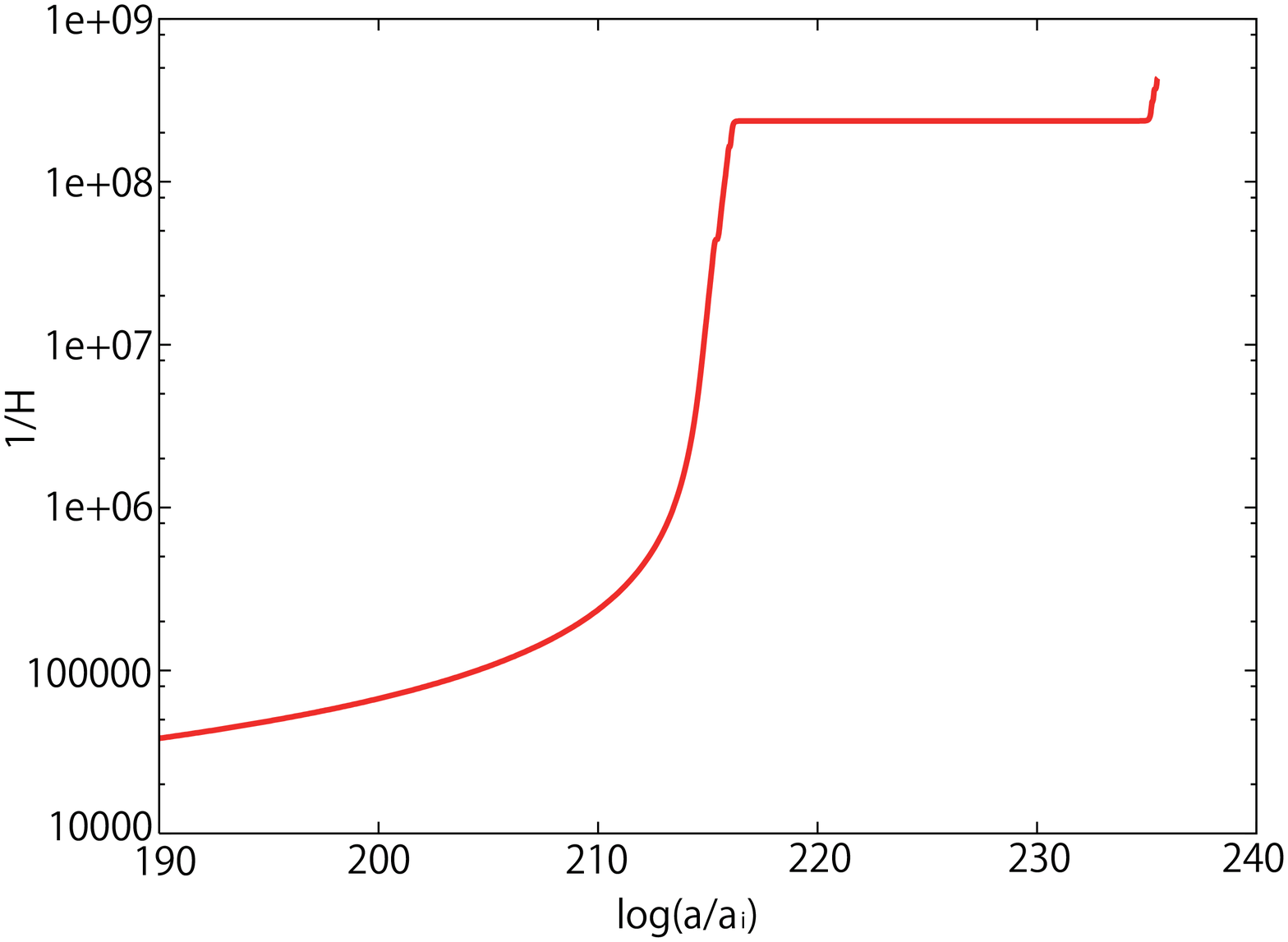}
		\includegraphics[width=.46\linewidth]{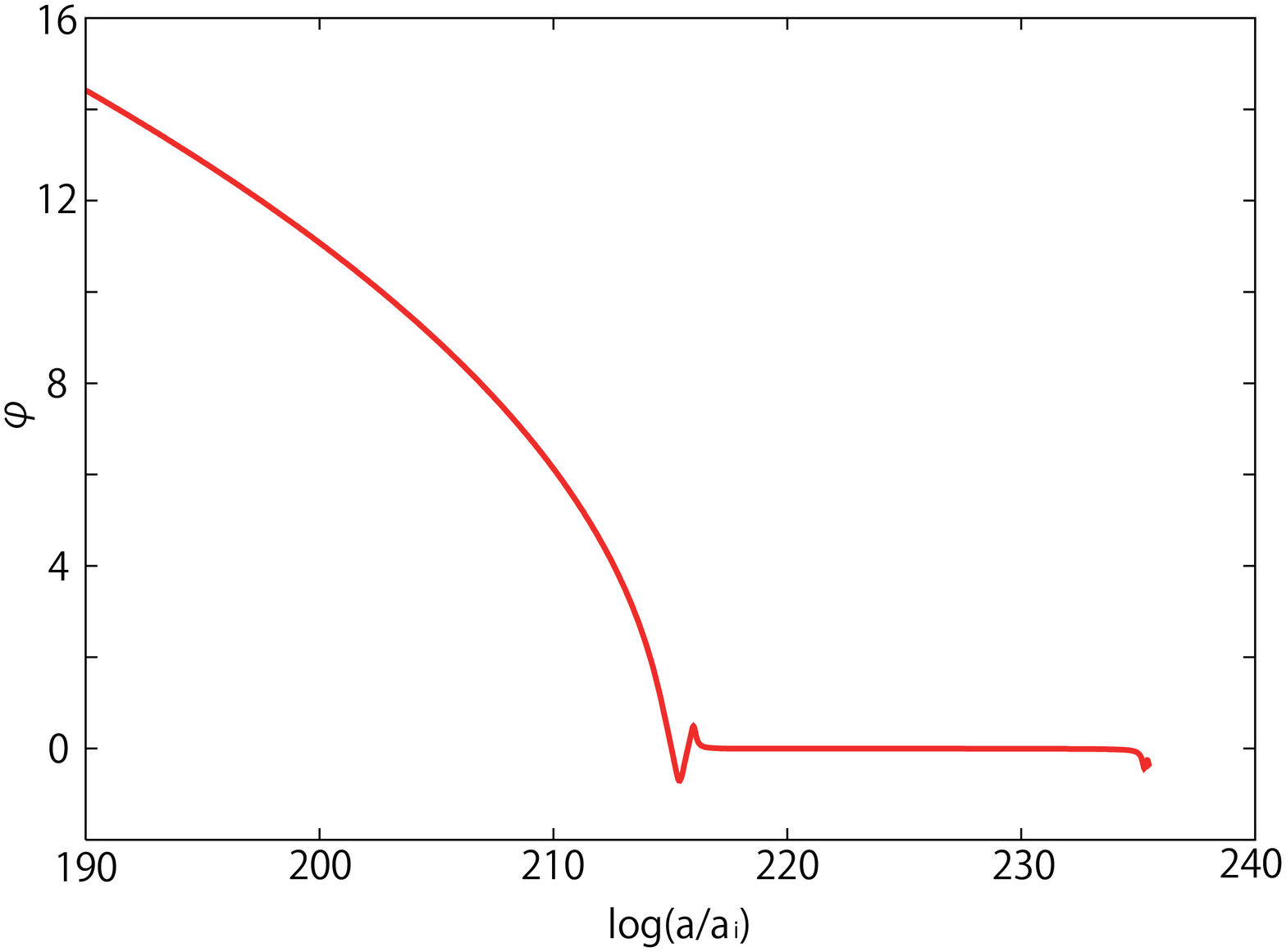}			
		\caption{The evolution of Hubble parameter(left) and the inflaton(right) with the values of the parameters $(\lambda,~v)=(5.4\times10^{-14},~0.355139M_{G})$. $a_i$ is a value of scale factor at the initial time. }
		\label{fig:bg}
	\end{figure}

In this model, inflation can occur twice \cite{yokoyama1999}. First, chaotic inflation occurs. After chaotic inflation, inflaton oscillates between the two minima of the potential. If the parameter $v$ is appropriately chosen, the inflaton moves slowly in the neighborhood of the origin after the oscillation and new inflation occurs. For example, the number of e-folds of new inflationay expansion, $N_{\text{new}}$, with $|\dot{H}|<H^2$ is larger than $10$ for $v=1.103M_{G}-1.132M_{G}$, and it satisfies $N_{\text{new}} \simg 60$ for $v=1.114M_{G}-1.122M_{G}$. According to the number of oscillation cycles, the parameters where new inflation occurs are distributed at certain intervals. In the above example, $\vphi$ settles to the positive potential minimum $\vphi=v$ if $v \ge 1.119M_{G}$ and to the negative potential minimum $\vphi=-v$ if $v \le 1.118M_{G}$ without oscillation. New inflation with $N_{\text{new}}>10$ occurs after a half cycles of oscillation for $v=0.35510M_{G}-0.35524M_{G}$ and after a cycle of oscillation for $v=0.266665M_{G}-0.26670M_{G}$ and so on. We show the evolution of Hubble parameter $H$ and the inflaton $\vphi$ with the values of the parameters $(\lambda,~v)=(5.4\times10^{-14},~0.355139M_{G})$ in Fig.~\ref{fig:bg}, which shows that the inflaton moves slowly in the neighborhood of the origin, and new inflation occurs. With these values of the parameters, $\vphi$ settles to the negative potential minimum after an oscillation.

\begin{figure}[h]
		\centering
		\includegraphics[width=.46\linewidth]{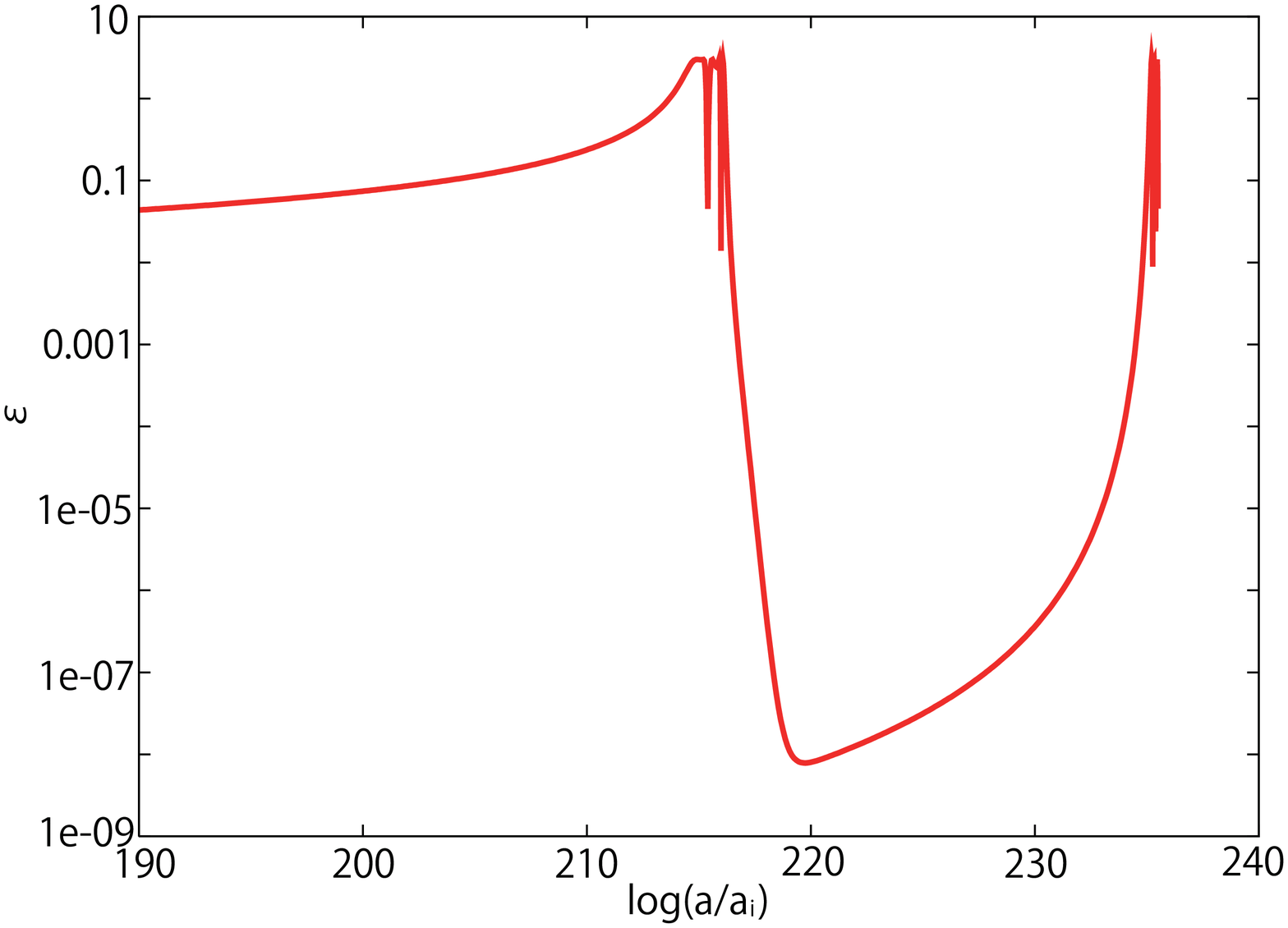}
		\includegraphics[width=.47\linewidth]{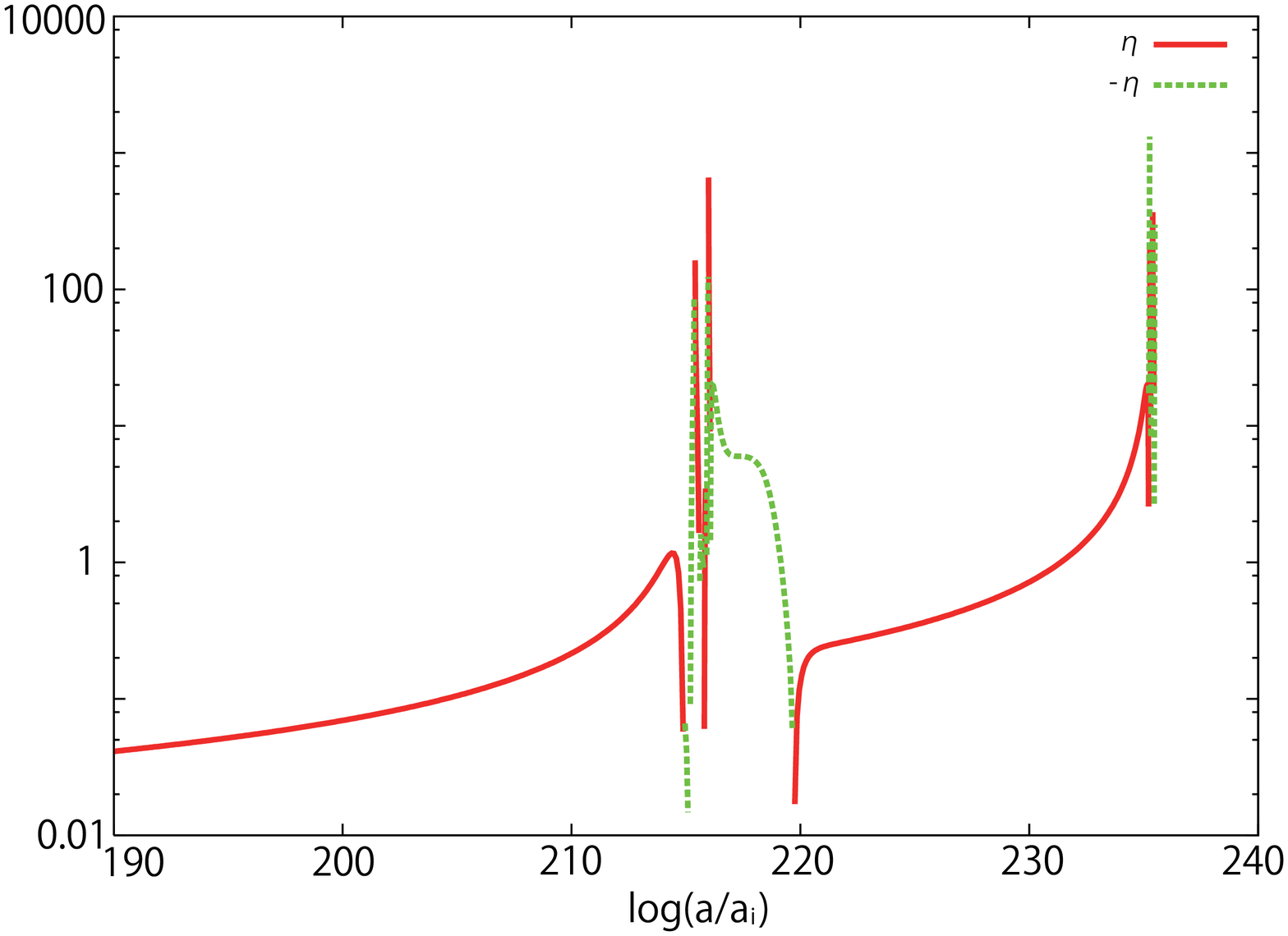}		
		\caption{The evolution of slow-roll parameters with the values of the parameters $(\lambda,~v)=(5.4\times10^{-14},~0.355139M_{G})$. }
		\label{fig:sr}
	\end{figure}

In describing the evolution of the inflaton during inflation, it is convenient to introduce the following Hubble slow-roll parameters:
	\begin{align}
		\eps &\equiv -\frac{\dot{H}}{H^2} = \frac{\dot{\vphi}^2}{2M_{G}^2H^2}, \label{eq:eps} \\
		\eta &\equiv \frac{\dot{\eps}}{H\eps}, \label{eq:eta}
	\end{align}
where dots denote differentiation with respect to the cosmic time $t$. We show the evolution of these slow-roll parameters in Fig.~\ref{fig:sr}. The values of the model parameters are the same as those employed in Fig.~\ref{fig:bg}. We can see that slow-roll conditions are not satisfied while the inflaton is oscillating between the two minima. As we see in the next section, the existence of this period is important for enhancement of curvature perturbation.

 \section{Enhancement of curvature perturbation}\label{sec:enhance}
  Curvature perturbation is enhanced in the models where slow-roll conditions are temporarily broken as stated in Ref.~\cite{leach2001}. In this section, we briefly describe the mechanism of the enhancement and give a power spectrum of curvature perturbation in the chaotic new inflation model.
  

	\subsection{Evolution of curvature perturbation}
 Curvature perturbation in the comoving gauge $\zeta$, in terms of which the amplitude of perturbation in the intrinsic spatial curvature of the comoving slicing $\mc{R}_c$ is written as
	\begin{equation}
		\mc{R}_{c} = \frac{4}{a^2}\nabla^2 \zeta,
	\end{equation}
evolves according to an equation
	\begin{equation}\label{eq:zetaev}
		\dif{2}{\zeta_{\mb{k}}}{N} + (3-\eps+\eta)\dif{}{\zeta_{\mb{k}}}{N} + \maru{\frac{k}{aH}}^2 \zeta_{\mb{k}} = 0,
	\end{equation}	
where $N$ is the number of e-folds and $\zeta_{\mb{k}}$ is the Fourier transform of $\zeta$:
	\begin{equation}
		\zeta_{\mb{k}} \equiv \int\!\mr{d}^3x~\zeta(t,\mb{x})e^{-i\mb{k}\cdot\mb{x}}.
	\end{equation}
~~In the slow-roll inflation regime, the coefficient of the second term in Eq.~(\ref{eq:zetaev}) is positive and therefore the solutions of Eq.~(\ref{eq:zetaev}) in the long-wavelength regime, where the last term is negligible, are a constant mode and a decaying mode. The time derivative $\mr{d}\zeta_{\mb{k}}/\mr{d}N$, which corresponds to the decaying mode, diminishes in proportional to $a^{-2}$. If we neglect the last term in Eq.~(\ref{eq:zetaev}), $\mr{d}\zeta_{\mb{k}}/\mr{d}N$ diminishes as $a^{-3}$. However, the last term decreases as $(k/aH)^2 \zeta_{\mb{k}} \propto a^{-2}$. As a result the second term in Eq.~(\ref{eq:zetaev}) soon becomes comparable to the last term in Eq.~(\ref{eq:zetaev}), and $\mr{d}\zeta_{\mb{k}}/\mr{d}N$ diminishes as $a^{-2}$. Therefore $\zeta_{\mb{k}}$ soon becomes constant after horizon crossing and the power spectrum of curvature perturbation is given by the squared amplitude of the vacuum fluctuation\footnote{In subhorizon scales, $\zeta_{\mb{k}}$ is given by the quantum fluctuation in the vacuum state. Therefore, $\zeta_k \pi_{\zeta_{\mb{k}}}$ has the minimum value $1$ under the uncertainty relation $\zeta_{\mb{k}} \pi_{\zeta_{\mb{k}}} \ge 1$. Here, $\pi_{\zeta_{\mb{k}}}$ is the momentum conjugate to $\zeta_{\mb{k}}$ and expressed as $2a^3 \eps M_G^2 \dot{\zeta}_{\mb{k}}^{\ast}$. Then, approximating $\dot{\zeta_{\mb{k}}} \sim (k/a)\zeta_{\mb{k}}$, we obtain the amplitude (\ref{eq:vf}).}
	\begin{equation}\label{eq:vf}
		|\zeta_{\mb{k}}|^2 = \frac{1}{2\eps M_G^2}\frac{k^{-1}}{a^2}
	\end{equation}
at the time the mode crossed the horizon:
	\begin{align}
		\mc{P}_{\zeta}(k) &= \left.\frac{k^3|\zeta_{\mb{k}}|^2}{2\pi^2}\right|_{k=aH} \nonumber \\
		&= \left.\frac{1}{\eps}\left(\frac{H}{2\pi M_G}\right)^2\right|_{k=aH}. \label{eq:srps}
	\end{align}	
 The power spectrum given by Eq.~(\ref{eq:srps}) is well approximated by the form $\mc{P}_{\zeta}(k) \propto k^{n-1}$ where $n-1=-2\eps-\eta$, namely, a nearly scale-invariant spectrum with small value of slow-roll parameters. Since the observed amplitude is small and single-field inflation models generically gives spectral index $n<1$ \cite{kinney2002}, this power spectrum leads to a small number of primordial black holes. On the other hand, the coefficient  of the second term in Eq.~(\ref{eq:zetaev}) can be negative in the model where slow-roll conditions are temporarily broken such as the chaotic new inflation model, in which we find $\eps \simeq 3$ and $\eta < 0$ near the end of the oscillatory phase so that $3-\eps+\eta<0$. In this case, Eq.~(\ref{eq:zetaev}) has a growing mode solution instead of the decaying mode solution outside the horizon. Therefore $\zeta_{\mb{k}}$ grows even after horizon crossing and its amplitude is enhanced. This enhancement of the perturbation enables even a single-field inflation model to produce a large number of PBHs.
 

 \subsection{Power spectrum of curvature perturbation in the chaotic new inflation model}
\begin{figure}[h]
		\centering
		\includegraphics[width=.7\linewidth]{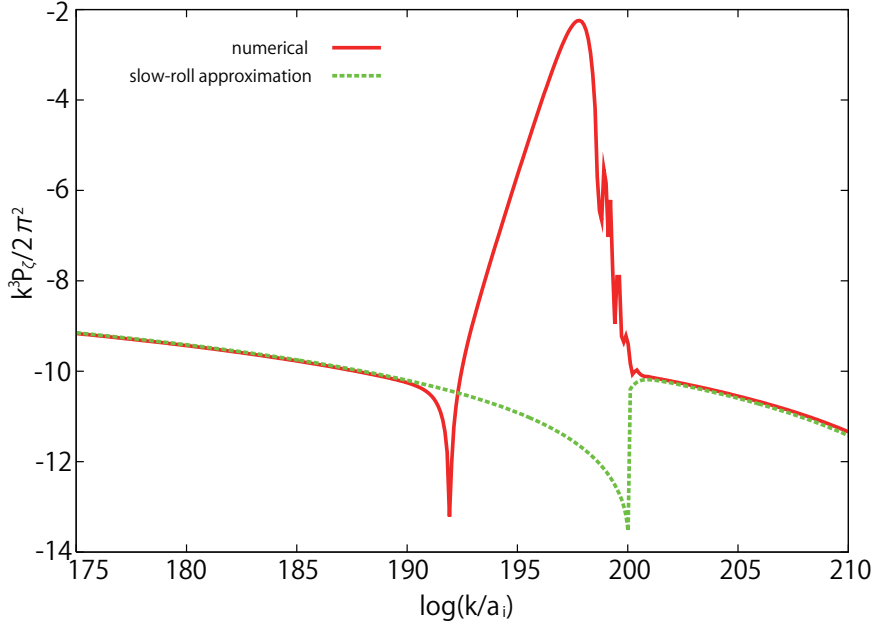}
		\caption{Power spectrum of curvature perturbation (solid line). This spectrum is calculated under the parameters $(\lambda,~v)=(5.4\times10^{-14},~0.355139M_{G})$. We show also a power spectrum estimated by using the formula (\ref{eq:srps}), which is used for a slow-roll inflation model (dashed line).}
		\label{fig:power}
	\end{figure}

 We have estimated the power spectrum of curvature perturbation in the chaotic new inflation model by solving Eq.~(\ref{eq:zetaev}) numerically. In Fig.~\ref{fig:power}, we show the power spectrum which is normalized to the amplitude observed by WMAP \cite{wmap5}, $2 \times 10^{-9}$ at $k=0.002/\mr{Mpc}$ by choosing $\lambda$ appropriately. The values of the model parameters are the same as those employed in Fig.~\ref{fig:bg} and Fig.~\ref{fig:sr}.
 
In Fig.~\ref{fig:power}, it is observed that the power spectrum deviates from the one estimated by using the slow-roll formula (\ref{eq:srps}) and the enhancement of the perturbation has occurred. The power spectrum has a peak with amplitude $\sim 6.2 \times 10^{-3}$ and results in formation of a large number of PBHs with mass corresponding to the scale of the peak. The scale of the peak corresponds to the scale which crossed the horizon near the end of chaotic inflation. The non-constant mode for the scales which crossed the horizon earlier is a decaying mode during slow-roll inflation, and becomes exponentially small as $a^{-2}$. Therefore, even if it turns to a growing mode temporarily after the first inflation, only the modes which left the horizon in the late stage of the first inflation are enhanced to a visible level. The scale and the amplitude of the peak vary with the parameter $v$. For increasing number of the oscillation cycles of inflaton, a larger amplitude is obtained. We investigate the abundance of PBHs resulted from the peaked spectrum in the following sections.


 \section{Abundance of primordial black holes}\label{sec:abundance}
 In this section, we give an expression of the PBH abundance resulting from a strongly peaked power spectrum.
 
We estimate the PBH abundance based on the Press-Schechter method \cite{press1973,liddlebook}. In this method, a PBH with mass greater than $M$ is formed when the perturbation which is smoothed on scale $R_M$ corresponding to $M$ exceeds the threshold $\zeta_{\mr{th}}$. The smoothed perturbation $\zeta_{R_M}$ is defined by
	\begin{equation}
		\zeta_{R_M}(\mb{x}) \equiv \int \mr{d}^3 x'~W(|\mb{x}'-\mb{x}|/{R_M})\zeta(\mb{x}'),
	\end{equation}	
where $W(x/R)$ is a window function. The fraction of the energy density of the Universe collapsing into PBHs with mass $M<M_{\mr{BH}}<M+\Delta M$ at the time they form is given by
	\begin{equation}\label{eq:beta}
		\beta(M;\Delta M) \equiv \frac{\rho_{\mr{BH}}(M;\Delta M)}{\rho_{\mr{tot}}} = -2\int_{M}^{M+\Delta M}\mr{d} M \int_{\zeta_{\text{th}}} \mr{d}\zeta~ \pdif{}{P_{R_M}}{M},
	\end{equation}
 where the prefactor 2 is due to Press-Schechter's prescription and $P_{R_M}$ is the probability distribution of $\zeta_{R_M}(\mb{x})$. $P_{R_M}$ is independent of $\mb{x}$ because of homogeneity of the universe. Therefore, we omit the argument $\mb{x}$ in the following. For the moment, we assume $\zeta_{\mb{k}}$ to be Gaussian distributed, which is the case to the lowest order of perturbation, and consider non-Gaussian correction in the next section to test the validity of this assumption. Then $P_{R_M}$ is Gaussian with variance
	\begin{equation}\label{eq:sigma}
		\sigma_{R_{M}}^2 \equiv \int\frac{\mr{d}k}{k}\wt{W}(kR)^2\mc{P}_{\zeta},
	\end{equation}
where $\wt{W}(kR)$ is the volume-normalized Fourier transform of the window function $W(x/R)$.

We estimate the PBH abundance, (\ref{eq:beta}), resulting from a strongly peaked power spectrum at the scale $k^{-1}=k_{\mr{peak}}^{-1}$. Since the power spectrum is strongly peaked we can approximate it to be $\delta$-function in Eq.~(\ref{eq:sigma}). This approximation is adequate, for $P_{R_M}$ is exponentially sensitive to $\sigma_{R_M}$ and a small difference in $\sigma_{R_M}$ leads to a large difference in $P_{R_M}$ so that only the value of $\sigma_R^2$ at the peak is important. Under this approximation and using a top-hat function as $\wt{W}(kR)$, we obtain $\mr{d} \sigma_{R_M}^2/\mr{d} M \sim \mc{P}_{\zeta}(k_{\mr{peak}})\delta(M-M_{\mr{peak}})$ where $M_{\mr{peak}}$ is the mass corresponding to the scale of the peak. Therefore, only PBHs whose mass is $M_{\mr{peak}}$ form. In this approximation, its abundance is given by
	\begin{equation}\label{eq:beta2}
		\beta(M_{\mr{peak}}) = \frac{1}{\sqrt{2\pi}}\int_{\hat{\zeta}_{\text{th}}} \mr{d}\hat{\zeta}~(\hat{\zeta}^2-1)e^{-\hat{\zeta}^2/2} \quad \maru{\hat{\zeta} \equiv \zeta/\sqrt{\mc{P}_{\zeta}(k_{\text{peak}})}},
	\end{equation}
where $\beta$ is independent of $\Delta M$ under the current approximation. In the peaked spectrum, the amplitude at the scale of the peak can be large while the amplitude at the observed scale is consistent with the observational value. Therefore, a large number of PBHs are produced from strongly peaked power spectrum.


 \section{Non-Gaussian correction to PBH abundance}\label{sec:ng}
 The amplitude of perturbation producing PBHs is so large that non-Gaussinity of curvature perturbation due to higher-order effects can be important. In the case of slow-roll inflation, this effect on the formation of PBHs has been studied in Ref.~\cite{pbhng} and shown to be negligibly small contrary to the intuitive expectation. However, in models where slow-roll conditions are temporarily violated such as the chaotic new inflation model, large non-Gaussinity arises \cite{chen2007} and can modify the PBH abundance obtained above. In this section, we investigate this possibility.


	\subsection{Three-point correlation functions}
  Here in order to estimate the effects of the deviation from Gaussian due to the higher-order interaction in the action, we first evaluate three-point correlation functions.
  
We sketch the derivation of the three-point correlation functions \cite{ng}. The three-point functions are estimated by perturbative expansion with respect to the interaction. For the estimation of the three-point correlation functions, the cubic terms of $\zeta$ \footnote{To be precise, {\it this} $\zeta$ is a generalization of $\zeta$ used in the linear perturbation theory \cite{ng}. At linear order, two $\zeta$ coincide with each other.} in the action give relevant interactions, which consist of
	\begin{align}
		S_3 &= \int\!\mr{d}t~L_3(\zeta,\dot{\zeta};t) \nm \\
	\begin{split}
		&= M_G^2 \int\!\mr{d}t \mr{d}^3 x  ~~[a\eps \zeta(\nabla\zeta)^2 + a^3\eps H^{-1}\dot{\zeta}^3 - 3a^3\eps \zeta\dot{\zeta}^2 \\
			&- \frac{1}{2a}(3\zeta-H^{-1}\dot{\zeta})(\nabla_i\nabla_j\psi\nabla_i\nabla_j\psi-\nabla^2\psi\nabla^2\psi) + 2a^{-1}\nabla_i\psi\nabla_i\zeta\nabla^2\psi], \label{eq:3rdaction}
	\end{split}
	\end{align} 
where $\psi$ is defined by
	\begin{equation}
		\psi \equiv -\frac{\zeta}{H}+a^2\eps \nabla^{-2}\dot{\zeta}.
	\end{equation}
 From these terms, the interaction Hamiltonian up to the third order reads
	\begin{align}
		H_I(\zeta_I,\pi_{\zeta I};t) &= - L_3(\zeta_I,\pi_{\zeta I}/(2a^3\eps);t) \nm \\
		&= - L_3(\zeta_I,\dot{\zeta_I};t) \nm \\
		&= M_G^2 \int\!\mr{d}^3 x  ~~[-a\eps \zeta_I(\nabla\zeta_I)^2 - a^3\eps H^{-1}\dot{\zeta_I}^3 + 3a^3\eps \zeta_I\dot{\zeta_I}^2 \nm \\
			&+ \frac{1}{2a}(3\zeta_I-H^{-1}\dot{\zeta_I})(\nabla_i\nabla_j\psi_I\nabla_i\nabla_j\psi_I-\nabla^2\psi_I\nabla^2\psi_I) - 2a^{-1}\nabla_i\psi_I\nabla_i\zeta_I\nabla^2\psi_I] \nm \\
		&= \frac{M_{G}^2 a^3 H^2}{( 2\pi )^6}\int\!\mr{d}^3 k_1\mr{d}^3 k_2\mr{d}^3 k_3~\delta^3(\mb{k}_1+\mb{k}_2+\mb{k}_3)~\left[ \mc{H}^{(1)}\zeta_{\mb{k}_1 I}\zeta_{\mb{k}_2 I}\zeta_{\mb{k}_3 I} \right. \nm \\
			& \qquad \left. + \mc{H}^{(2)}\zeta_{\mb{k}_1 I}\zeta_{\mb{k}_2 I}\frac{\mr{d} \zeta_{\mb{k}_3 I}}{H \mr{d}t} + \mc{H}^{(3)}\frac{\mr{d} \zeta_{\mb{k}_1 I}}{H \mr{d}t}\frac{\mr{d} \zeta_{\mb{k}_2 I}}{H \mr{d}t}\zeta_{\mb{k}_3 I} + \mc{H}^{(4)}\frac{\mr{d} \zeta_{\mb{k}_1 I}}{H \mr{d}t}\frac{\mr{d} \zeta_{\mb{k}_2 I}}{H \mr{d}t}\frac{\mr{d} \zeta_{\mb{k}_3 I}}{H \mr{d}t} \right], \label{eq:interaction}
	\end{align}
where $\pi_{\zeta}$ is the momentum conjugate to $\zeta$ and the variables with subscript $I$ denote variables in the interaction picture. Coefficients $\mc{H}^{(i)}~(i=1,2,3,4)$ are given by
	\begin{align}
		\mc{H}^{(1)} &= \eps\bar{k}^2\cos \theta - \frac{1}{6}\bar{k}^4\sin^2\theta, \label{eq:h1} \\
		\mc{H}^{(2)} &= -\eps \bar{k}^2 (\sin^2 \theta/z + \cos \theta) + \frac{\bar{k}^4}{2}\sin^2 \theta, \label{eq:h2} \\
		\mc{H}^{(3)} &= 3\eps + \frac{\eps^2}{2}(\sin^2\theta + 2z\cos\theta) + \frac{\eps}{2}\bar{k}^2(z-2\cos\theta)\sin^2\theta, \label{eq:h3} \\
		\mc{H}^{(4)} &= -\eps - \frac{\eps^2}{2}\sin^2\theta, \label{eq:h4}
	\end{align}
where we have defined
	\begin{equation}\label{eq:mod}
		\bar{k} \equiv \frac{\sqrt{k_1k_2}}{aH}, \quad \cos \theta \equiv (\mb{k}_1 \cdot \mb{k}_2)/k_1k_2, \quad z \equiv k_3^2/k_1k_2,
	\end{equation}
for simplicity of expression. For an equilateral triangle, $k_1=k_2=k_3=k$, the values of these quantities are
	\[
		\bar{k} = \frac{k}{aH}, \quad \cos \theta = -\frac{1}{2}, \quad z =1.
	\]
~~In calculating the three-point correlation functions, the in-in formalism \cite{in-in} is used since we want to calculate an expectation value with respect to the vacuum state at $t \to -\infty$:
	\begin{equation}\label{eq:int3pt}
	\begin{split}
		\kaku{\zeta_{\mb{k}_1}(t)\zeta_{\mb{k}_2}(t)\zeta_{\mb{k}_3}(t)} = &\kaku{{U_I}^{-1}(t,t_0)\zeta_{\mb{k}_1 I}(t)\zeta_{\mb{k}_2 I}(t)\zeta_{\mb{k}_3 I}(t)U_I(t,t_0)}. \\ & \qquad \quad \left( U_I(t,t_0) \equiv Te^{-i\int_{(1-i\eps_0)t_0}^{t}\!\mr{d}t~H_I(\zeta_I(t),\pi_{\zeta I}(t);t)} \right)
	\end{split}
	\end{equation}
where $\eps_0$ is a positive infinitesimal constant\footnote{In an abuse of language, we use the same symbol $\zeta_{\mb{k}}$ for denoting a quantized variable as a classical one.}. To leading order of $H_I$, the three-point correlation function is given by 
	\begin{align}
		\kaku{{U_I}^{-1}\zeta_{\mb{k}_1 I}(t)\zeta_{\mb{k}_2 I}(t)\zeta_{\mb{k}_3 I}(t)U_I} &= -i\int_{t_0}^{t}\!\mr{d}t'\kaku{[\zeta_{\mb{k}_1 I}(t)\zeta_{\mb{k}_2 I}(t)\zeta_{\mb{k}_3 I}(t),H_I(\zeta_{\mb{k} I},\pi_{\zeta_{\mb{k}} I};t')]} \nm \\
		&= 2\int_{t_0}^{t}\!\mr{d}t' \mr{Im}\left(\kaku{\zeta_{\mb{k}_1 I}(t)\zeta_{\mb{k}_2 I}(t)\zeta_{\mb{k}_3 I}(t)H_I(\zeta_{\mb{k} I},\pi_{\zeta_{\mb{k}} I};t')} \right), \label{eq:3pt}
	\end{align}
where the $t'$ integration contour is deformed so that both bra and ket are projected on to the vacuum state at $t \to -\infty$. According to the behavior of $\zeta_{\mb{k} I}$, the integrals in Eq.~(\ref{eq:3pt}) can be splitted into three parts, an integral over the region inside the horizon, the region around horizon crossing and the region outside the horizon. In the former two parts the deviation from the slow-roll inflation models is small, since the third term in Eq.~(\ref{eq:zetaev}) is dominant in these parts. The deviation arises in the last part, because $\zeta_{\mb{k}}$ grows outside the horizon in the chaotic new inflation model and not in the slow-roll inflation models. In the slow-roll inflation models, the contribution from the part outside the horizon is negligible since each term in the interaction (\ref{eq:interaction}) includes $\dot{\zeta}_I$ or $k/aH$ which have small values outside the horizon and, furthermore, the commutator of $\zeta_{\mb{k}}$'s or its time derivatives vanish as $a^{-\nu}~(\nu \ge 2)$ \footnote{If $\zeta_{\mb{k} I}(t)$ is constant $\zeta_{\mb{k} I}(t)=\zeta_{\mb{k} I}^{\text{const}}$, the three-point functions vanishes because a term $\zeta_{\mb{k}_1 I}(t)\zeta_{\mb{k}_2 I}(t)\zeta_{\mb{k}_3 I}(t)\zeta^{\ast}_{\mb{k}_1 I}(t')\zeta^{\ast}_{\mb{k}_2 I}(t')\zeta^{\ast}_{\mb{k}_3 I}(t')$ becomes a real number $|\zeta_{\mb{k}_1 I}^{\text{cosnt}}\zeta_{\mb{k}_2 I}^{\text{const}}\zeta_{\mb{k}_3 I}^{\text{const}}|^2$ in this case. Then the leading term of the commutator of $\zeta_{\mb{k}}$'s have a decaying mode in $\zeta_{\mb{k}}$'s, and decreases as $a^{-2}$. With time derivatives of $\zeta_{\mb{k}}$'s, the commutators decreases more rapidly.}. The three-point correlation functions are suppressed by the slow-roll parameters estimated at horizon crossing \cite{ng}. In contrast, in the chaotic new inflation model, $\dot{\zeta}_I$ has a non-negligible value outside the horizon and the integrals over the region outside the horizon contributes to the three-point correlation functions. The reason for obtaining large non-Gaussianity is different from that in Ref.~\cite{chen2007}. In the chaotic new inflation model it is the growth of the perturbation outside the horizon, while in Ref.~\cite{chen2007} it is the characteristic behavior of the slow roll parameters near or inside the horizon.


	\subsection{Correction to PBH abundance from three-point correlation functions}
  In the following, we give an expression of the PBH abundance resulting from the perturbation with the three-point correlation functions.
  
The probability distribution $P_{R_M}$ can be expressed as 
	\begin{equation}
		P_{R_M}(\zeta_{R_M}) = \frac{1}{2\pi}\int \mr{d} \eta_{R_M}~\Phi(\eta_{R_M})e^{-i\eta_{R_M}\zeta_{R_M}},
	\end{equation}
where $\Phi$ is defined by
	\begin{equation}
		\Phi(\eta_{R_M}) \equiv \kaku{e^{i\eta_{R_M}\zeta_{R_M}}}.
	\end{equation}
 $\Phi$ can be expanded by cumulants of $\zeta_{R_M}$, $\kaku{\zeta_{R_M}^n}_c$:
	\begin{equation}\label{eq:phi}
		\Phi(\eta_{R_M}) = \exp\left( \sum_{m=0}^{\infty} \frac{(i\eta_{R_M})^m}{m!}\kaku{\zeta_{R_M}^m}_c \right).
	\end{equation}
 The cumulants of $\zeta_{R_M}$ can be expressed by the connected part of the correlation functions of $\zeta_{\mb{k}}$. With the formula (\ref{eq:phi}), we can estimate the correction to $P_{R_M}$ from higher-order correlation functions.
 
The correction from the three-point correlation functions can be calculated, retaining terms up to $m=3$ in Eq.~(\ref{eq:phi}). In the case $|\kaku{\zeta_{R_M}^3}_c|$ is much smaller than $\sigma_{R_M}^3$, we can give a concrete expression of the corrected PBH abundance. In this case, we can expand $\Phi$ with respect to $J$ which is defined by
	\begin{equation}
		J \equiv \frac{1}{6} \frac{\kaku{\zeta_{R_M}^3}_c}{\sigma_{R_M}^3},
	\end{equation}
and get
	\begin{equation}\label{eq:p}
		P_{R_M}(\zeta_{R_M}) = \frac{1}{\sqrt{2\pi\sigma_{R_M}^2}}\left[ 1+\left(3\frac{\zeta_{R_M}}{\sigma_{R_M}}-\frac{\zeta_{R_M}^3}{\sigma_{R_M}^3}\right)J \right]\exp\maru{-\frac{\zeta_{R_M}^2}{2\sigma_{R_M}^2}}.
	\end{equation}
 $J$ can be expressed by the three-point correlation functions of $\zeta_{\mb{k}}$ as
	\begin{equation}\label{eq:j}
		J = \frac{1}{6\sigma_{R_M}^3}\int\!\mr{d}^3k_1\int\!\mr{d}^3k_2\int\!\mr{d}^3k_3~\wt{W}(k_1R)\wt{W}(k_2R)\wt{W}(k_3R)\kaku{\zeta_{\mb{k}_1}\zeta_{\mb{k}_2}\zeta_{\mb{k}_3}}.
	\end{equation}
Substituting the probability distribution $P_{R_M}$ given by Eq.~(\ref{eq:p}) to Eq.~(\ref{eq:beta}), we get the PBH abundance including the correction from the three-point correlation functions.

\begin{figure}[h]
		\centering
		\includegraphics[width=.7\linewidth]{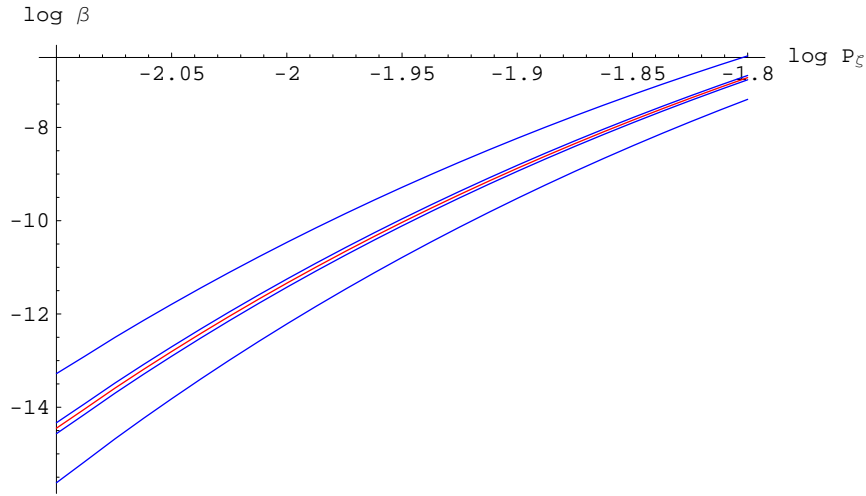}
		\caption{PBH abundance with some values of $J_{\text{peak}}$. The values of $J_{\text{peak}}$ are $-0.1,-0.01,0,0.01,0.1$ from below.}
		\label{fig:j}
	\end{figure}

Because of homogeneity of the universe, the three-point correlation functions of $\zeta_{\mb{k}}$ can be written as
	\begin{equation}
		\kaku{\zeta_{\mb{k}_1}(t)\zeta_{\mb{k}_2}(t)\zeta_{\mb{k}_3}(t)}_c = A(\mb{k}_1,\mb{k}_2, \mb{k}_3)\delta^3(\mb{k}_1+\mb{k}_2+\mb{k}_3).
	\end{equation}
The isotropy of the universe guarantees that $A(\mb{k}_1,\mb{k}_2, \mb{k}_3)$ is in fact a function of ``shape'' and ``size'' of triangle spanned by $\mb{k}_{i}~(i=1,2,3)$ such as the quantities (\ref{eq:mod}) and independent of direction of the triangle. In the chaotic new inflation model, $\zeta_{\mb{k
}}$ grows most significantly at $k=k_{\text{peak}}$ and $A(\mb{k}_1,\mb{k}_2, \mb{k}_3)$ has the largest value at $k_1=k_2=k_3=k_{\text{peak}}$. Therefore, as before, we can take approximation $\mr{d} \kaku{\zeta_{R_M}^3}_c/\mr{d} M \sim 4\pi k_{\text{peak}}^6 A_{\text{peak}} \delta(M-M_{\text{peak}})$, where $A_{\text{peak}}$ represents $A(\mb{k}_1,\mb{k}_2, \mb{k}_3)$ estimated at $k_1=k_2=k_3=k_{\text{peak}}$. Under this approximation, the PBH abundance is given by
	\begin{align}
		\beta(M_{\text{peak}}) &= \frac{1}{\sqrt{2\pi}}\int_{\hat{\zeta}_{\text{th}}} \mr{d}\hat{\zeta}~\left[(\hat{\zeta}^2-1) - (\hat{\zeta}^5-8\hat{\zeta}^3+9\hat{\zeta})J_{\text{peak}}\right]e^{-\hat{\zeta}^2/2} \label{eq:beta3} \\
	&\qquad \qquad \qquad \qquad \qquad \qquad \qquad \qquad \maru{\hat{\zeta} \equiv \zeta/\sqrt{\mc{P}_{\zeta}(k_{\text{peak}})}}, \nm
	\end{align}
where
	\begin{equation}\label{eq:jpeak}
		J_{\text{peak}} \equiv \frac{2\pi}{3} \frac{ k_{\text{peak}}^6 A_{\text{peak}}}{\mc{P}_{\zeta}(k_{\text{peak}})^{3/2}}.
	\end{equation}
The PBH abundance with some values of $J_{\text{peak}}$ is shown in Fig.~\ref{fig:j}.

	\subsection{Estimation of correction to PBH abundance in chaotic new inflation model}
  With the formulae obtained in the previous subsections, we can estimate the correction to the PBH abundance from the three-point correlation functions in the chaotic new inflation model. We have carried out the integration in Eq.~(\ref{eq:3pt}) numerically over the region outside the horizon, where the deviation from the slow-roll inflation models arises. In the other regions the contributions are almost the same as those in the slow-roll inflation models and do not modify the PBH abundance relevantly. Since the calculation is done only outside the horizon, we do not have the problem in implementing the $i \epsilon$ prescription numerically.
  
For the values of the parameters $(\lambda,~v)=(5.4\times10^{-14},~0.355139M_{G})$ with which the amplitude at the peak has the values for producing relevant number of PBHs, $\sim 6.2 \times 10^{-3}$, the value of $A_{\text{peak}}$ is estimated to be
	\begin{equation}
		k_{\text{peak}}^6 A_{\text{peak}} \sim 10^{-9}.
	\end{equation}
On the other hand, the value obtained by estimating the contribution from the region around horizon crossing as done in slow-roll inflation models is calculated to be \cite{ng}
	\begin{equation}
		k_{\text{peak}}^6 A_{\text{peak}} = 48\pi^7 f_{\mr{NL}}\mc{P}_{\zeta}^2 \sim 10^{-16},
	\end{equation}
where the value of $\mc{P}_{\zeta}$ is estimated at horizon crossing, $\sim 10^{-11}$ (see Fig.~\ref{fig:power}). $f_{\mr{NL}}$ is an estimator usually used for parameterizing the size of non-Gaussianity observed in the CMB \cite{ng} and has the value of the order of the slow-roll parameters estimated at horizon crossing, $\sim 10^{-1}$ (see Fig.~\ref{fig:sr}). 

We get larger three-point correlation functions than those without the enhancement of the perturbation. However, the parameter $J_{\text{peak}}$ which is an estimator for the correction to the PBH abundance from the three-point correlation functions is estimated to be
	\begin{equation}
		J_{\text{peak}} \sim 10^{-6}.
	\end{equation}
This value is too small to modify the PBH abundance relevantly. Though large three-point correlation functions are obtained, since a denominator of the estimator is also large, only a small correction is obtained.\\
  The correction to the PBH abundance from $N$-point correlation functions appears with a factor $\mc{P}_{\zeta}^{-N/2}$. Therefore the corrections are expected to be small as the correction from three point functions are. We can therefore use the expression (\ref{eq:beta2}) for the PBH abundance safely. 


 \section{Parameter search}\label{sec:search}
  In \S \ref{sec:enhance}, we have shown the curvature perturbation $\zeta$ is enhanced in the chaotic new inflation model. In this section, we calculate the PBH abundance using Eq.~(\ref{eq:beta2}) with various values of the parameter $v$ and search the parameter with which relevant number of PBHs can be produced. For each $v$, the parameter $\lambda$ is fixed by the power spectrum of $\zeta$ observed by WMAP, $\mc{P}_{\zeta}=2 \times 10^{-9}$ at $k=0.002/\mr{Mpc}$. In the linear perturbation theory, the power spectrum $\mc{P}_{\zeta}(k)$ scales as\footnote{The equations in the linear perturbation theory and the initial conditions for $V'=\wt{\lambda} V$ can be written as those for $V$ by defining $(N',H',\vphi',k',\zeta') \equiv (N,\wt{\lambda}^{-1/2}H,\vphi,\wt{\lambda}^{-1/2}k,\wt{\lambda}^{1/4}\zeta)$. Then, the relation (\ref{eq:scale}) is obtained.}
	\begin{equation}\label{eq:scale}
		\mc{P}_{\zeta}(k) \to {\mc{P}_{\zeta}'}(k)=\wt{\lambda}\mc{P}_{\zeta}(\wt{\lambda}^{-1/2}k) \quad \text{for} \quad V \to V'=\wt{\lambda} V.
	\end{equation}
  Since the horizon mass at the matter-radiation equality is $\sim 10^{17}M_{\odot}$, PBHs that can be the origin of intermediate mass black holes or dark matter are produced in the radiation-dominated epoch. In the radiation-dominated epoch, the threshold value of the density perturbation $\delta_{\mr{th}}$ is given by $1/3$ \cite{pbh,carr2003}. Then, we find the corresponding value of the curvature perturbation $\zeta_{\text{th}}=0.75$ with the help of the formula in the linear perturbation theory, $\delta_{k=aH}=4\zeta_{k=aH}/9$ \cite{liddlebook}. A similar value has been obtained by numerical calculation in Ref.~\cite{shibata1999}. Equation (\ref{eq:bhmass}) relates the mass of PBH produced in the radiation-dominated epoch to the scale of the perturbation as
	\begin{equation}\label{eq:mr}
		M = 5.4~\mr{g}~\left(\frac{10^{16}~\mr{GeV}}{V_{\ast}^{1/4}}\right)^2\!\left(\frac{a_{\ast}R_M}{H_{\ast}^{-1}}\right)^{2},
	\end{equation}
where the variables with subscript ``$\ast$" denote those estimated at arbitrary time in the radiation-dominated epoch. We assume the period of reheating is negligible and have chosen the time at which inflation finished as time at which the variables with subscript ``$\ast$" are estimated. We write these variables with subscript ``end". 

\begin{figure}[h]
		\centering
		\includegraphics[width=.45\linewidth]{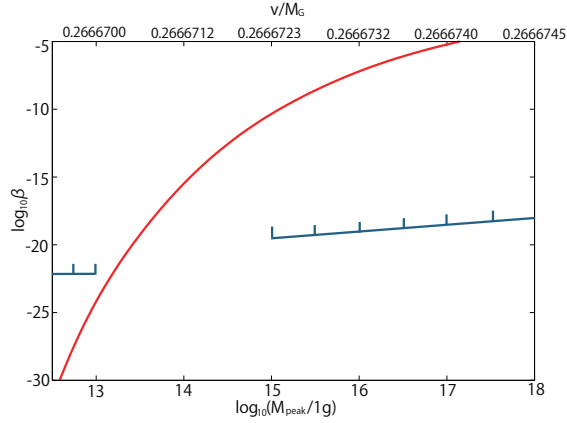}
	\caption{PBH abundance $\beta$ as a function of mass corresponding to the scale of the peak $M_{\text{peak}}$ associated with the values of parameter $v=0.2666694M_{G}-0.2666745M_{G}$ (solid line). The constraints on PBH abundance are also depicted.}
		\label{fig:beta5}
	\end{figure}

\begin{figure}[h]
		\centering
		\includegraphics[width=.45\linewidth]{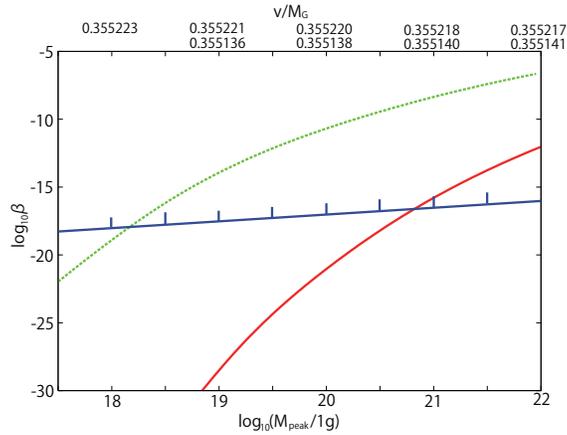}
		\caption{PBH abundance $\beta$ as a function of mass corresponding to the scale of the peak $M_{\text{peak}}$ associated with the values of parameter $v=0.355136M_{G}-0.355141M_{G}$ (solid line) and $v=0.355217M_{G}-0.355223M_{G}$(dashed line). The constraint on PBH abundance is also depicted.}
		\label{fig:beta7}
	\end{figure}

We have found that a relevant number of PBHs can be produced around $v \simeq 0.26667M_{G}$, $v \simeq 0.35514M_{G}$ and $v \simeq 0.35522M_{G}$. In these values of $v$, $V_{\text{end}}^{1/4} \sim (\lambda v^4)^{1/4} \sim 10^{14}~\mr{GeV}$ and  Eq.~(\ref{eq:mr}) gives
	\begin{equation}\label{eq:mr2}
		M \sim 10^{4}~\mr{g}~\left(\frac{R_M}{R_{\text{end}}}\right)^{2},
	\end{equation}	
where $R_{\text{end}} \equiv 1/(a_{\text{end}}H_{\text{end}})$, which is the comoving scale crossing the horizon at the end of the entire inflation. The PBH abundance $\beta$ is plotted as a function of the mass corresponding to the scale of the peak $M_{\text{peak}}$ in Figs.~\ref{fig:beta5} and \ref{fig:beta7} with the observational constraints. The constraint on PBH abundance with mass above $10^{15}~\mr{g}$ is obtained from the condition that the abundance of PBHs is less than that of matter today. The other constraints are obtained from the consistency with nucleosynthesis and $\gamma$-ray observation. In Fig.~\ref{fig:beta7}, we can observe that PBHs with mass $6 \times 10^{20}~\mr{g}$ can constitute a large part of matter in the universe. Though PBHs with some range of mass are observationally excluded to be the dominant component of dark matter \cite{pbhdm}, there are no strong constraints in mass range $10^{20}~\mr{g}-10^{26}~\mr{g}~(10^{-13}M_{\odot}-10^{-7}M_{\odot})$. The mass $6 \times 10^{20}~\mr{g}$ are in this range. Therefore, we can obtain PBHs that can be an origin of dark matter in the chaotic new inflation model.


Finally, we estimate scalar spectral index $n$ and tensor-to-scalar ratio $r$ and check the consistency with the observational data \cite{wmap5}:
	\begin{equation}
		n = 0.986 \pm {0.022}, \quad r<0.43~(95\% \text{CL}) \quad \text{at} \quad k=0.002/\mr{Mpc}.
	\end{equation}
 In each parameter regions, scalar spectral index $n$ and tensor-to-scalar ratio $r$ is estimated to be
	\begin{align*}
		n = 0.939-0.931, \quad r=0.33-0.37 \qquad \text{for} \quad v = 0.266669M_{G}-0.266674M_{G}, \\
		n = 0.928-0.922, \quad r=0.38-0.42 \qquad \text{for} \quad v = 0.355136M_{G}-0.355141M_{G}, \\
		n = 0.931-0.924, \quad r=0.37-0.41 \qquad \text{for} \quad v = 0.355223M_{G}-0.355217M_{G},
	\end{align*}
especially
	\[
		n=0.925, \quad r=0.40 \qquad \text{for} \quad v = 0.355140M_{G}~(M_{\text{peak}}=6 \times 10^{20}~\mr{g}).
	\]
 At the first parameter region, $n$ and $r$ are consistent with 3-year WMAP data at the confidence level of $99.9\%$, but lie out of the preferred range of 5-year data. At the other regions,  $n$ and $r$ also lie out of the preferred range of 5-year data.
 

 \section{Discussion}\label{sec:conclusion}
 In this paper, we have shown that the enhancement of the perturbation occurs and a large number of PBHs are produced in the chaotic new inflation model. A growing mode plays an important role in the enhancement. Due to the growing mode, there is a contribution to the three-point functions from the region outside the horizon. We have estimated this contribution and its effect on the PBH abundance. As a result, We have found that non-Gaussian correction to the PBH abundance due to higher-order effect is small. Further, we have calculated the PBH abundance with various values of the parameter $v$. We have obtained the relevant number of PBHs around $v \simeq 0.26667M_{G}$, $v \simeq 0.35514M_{G}$ and $v \simeq 0.35522M_{G}$. In the second parameter region, the produced PBHs can constitute a large part of dark matter in the Universe. In the first parameter region, though the produced PBHs cannot be observed, we can observe the enhancement of the perturbation with gravitational waves generated from scalar perturbations through non-linear couplings \cite{ananda2007}. The scale of the peak which gives mass $M=1.4 \times 10^{13}~\mr{g}$ corresponds to GW frequency\footnote{The power spectrum of the induced GW have the peak at $k_{\mr{peak}}/\sqrt{3}\pi$ \cite{ananda2007}.} $\sim 100~\mr{Hz}$, and power spectrum for the induced GW is estimated to be\footnote{In Ref.~\cite{ananda2007}, the energy density of the induced GW is calculated to be $\Omega_{\mr{GW}} \sim 10^{-17} \mc{A}^4$ where $\mc{A}$ is the amplitude at the scale of the peak relative to the observed amplitude. At $M=1.4 \times 10^{13}~\mr{g}$, $\mc{A}$ is estimated to be $\sim 10^{3}$, and $\Omega_{\mr{GW}} \sim 10^{-5}$. This value is comparable to the nucleosynthesis bound $\Omega_{\mr{GW}} \sim 10^{-5}$ \cite{maggiore2000}. However, the amplitude of the scalar perturbations is so large that the effects neglected in Ref.~\cite{ananda2007} such as backreactions to the scalar perturbations from the tensor perturbations should be properly taking into account.} $\sim \mc{P}_{\zeta}^2 \sim 10^{-4}$. Therefore, the induced GW can be detected by detectors such as GEO600 \cite{geo}, LIGO \cite{ligo}, TAMA \cite{tama}, VIRGO \cite{virgo}.

In both parameter regions,  however, the values of spectral index and tensor-to-scalar ratio are not in the preferred region of the observational data. This is because, in the chaotic new inflation model, the observed perturbations are produced at the chaotic inflationary epoch with a potential close to the quartic one, which is not preferred by the observational data. In this paper we adopted the Coleman-Weinberg potential as a simple model to drive inflationary dynamics without specifying particle physics background. In reality, we expect supergravity corrections to the scalar potential is important, especially for $\vphi>M_{G}$. If such corrections modifies the potential in an appropriate way, this scenario could be a feasible one. Another way to rescue it is to adopt a smaller self coupling $\lambda$ using the scaling law (\ref{eq:scale}), so that it predicts smaller amplitudes of both scalar and tensor fluctuations on large scales observed by CMB etc., assuming that the observed scalar perturbations on these scales were created through a different mechanism, say, curvaton \cite{curvaton}, or modulated rehating \cite{mreheat}. Since the peak we have found is so prominent that one could produce observable amount of PBHs even in such a case with smaller $\lambda$.

\section*{Acknowledgement}

This work was partially supported by
JSPS Grant-in-Aid for Scientific Research Nos.~16340076 and 19340054.

\end{document}

%% file: mydefines.tex
\def\mr{\mathrm}
\def\mb{\boldsymbol}
\def\mc{\mathcal}
\def\nm{\nonumber}
\def\vphi{\varphi}
\def\eps{\epsilon}
\def\pl{\partial}

\def\wt{\widetilde}

\def\simg{\hspace{0.3em}\raisebox{0.4ex}{$>$}\hspace{-0.75em}\raisebox{-0.7ex}{$\sim$}\hspace{0.3em}}

\newcommand{\dif}[3]{\frac{\mr{d}^{#1} #2}{\mr{d} {#3}^{#1}}}
\newcommand{\pdif}[3]{\frac{{\pl}^{#1} #2}{{\pl} {#3}^{#1}}}

\newcommand{\kaku}[1]{\langle #1 \rangle}
\newcommand{\maru}[1]{\left( #1 \right)}